\documentclass[11pt]{article}
\usepackage{graphicx}
\begin{document}
\baselineskip=18pt
\DeclareGraphicsExtensions{.pdf,.png,.jpg}
\begin{center}

{\huge Steiner trees and spanning trees in six-pin soap films}

\vskip .5cm

{\bf Prasun Dutta\footnote{Corresponding author: 
Phone: +91-3222-281645, Fax: +91-3222-255303,
email: prasun@phy.iitkgp.ernet.in}, 
S. Pratik Khastgir and 
Anushree Roy}

{\it Department of
Physics and Meteorology, Indian Institute of Technology,}

{\it Kharagpur 721302, INDIA}

\end{center}
\begin{abstract}
We have studied the Steiner tree problem using six-pin soap films
in detail. We extend the existing  method of experimental 
realisation of Steiner trees
in $n$-terminal problem through soap films to observe new non-minimal
Steiner trees. We also produced spanning tree
configurations for the first time by our method.
Experimentally, by varying the pin diameter, we have
achieved these new stable soap film configurations.
A new algorithm is presented for creating  
these Steiner trees theoretically.
  Exact lengths of these Steiner tree configurations are calculated
using a geometrical method. 
An exact two-parameter empirical formula is proposed for 
estimating the lengths of these soap
film configurations in six-pin soap film problem.
\end{abstract}
{PACS: 45.10 Db}

{\it Keywords:} Steiner trees, Spanning trees, Soap films
\def\d{{\mathrm{d}}}

\section{Introduction}
The problem of finding minimum, local as well as absolute, path
lengths joining given points (or terminals) on a plane is the
well-known Steiner problem \cite{Courant}. The $n$-terminal 
generalised Steiner problem is yet unsolved. In the present note we
have studied the theoretical construction and experimental realisation
of {\it Steiner trees}  and {\it spanning trees} in a
6-terminal problem. To distinguish a 
{\it Steiner tree}  from a {\it spanning tree} we make ourselves
familiar with the basic terminology of elementary graph theory in the 
following. 

A {\it graph} consists of a finite number of {\it points(terminals or 
vertices)}  together with some or all pair of points joined with 
{\it lines(edges or branches)} \cite{Harary}. In what follows we 
shall assume all
the points on a plane and all the lines as straight.
 A {\it path} connects a pair of points by a line or many lines
 through the other points. In a {\it connected graph} there
 exists at least a path between every pair of points.
A {\it complete graph} $K_p$ has every pair of its $p$ points
connected by a line. A {\it subgraph} has all its points and lines in
the graph. A {\it spanning subgraph} is a subgraph containg all the
points of graph. A {\it tree} is a connected graph and has no cycles
formed by its lines(or branches). A {\it spanning tree} is a spanning
subgraph without any cycles in it. In the following we shall be
interested in the spanning trees of $K_p$ in general and $K_6$ in
particular.
 
\begin{figure}
\includegraphics[width=6.5in]{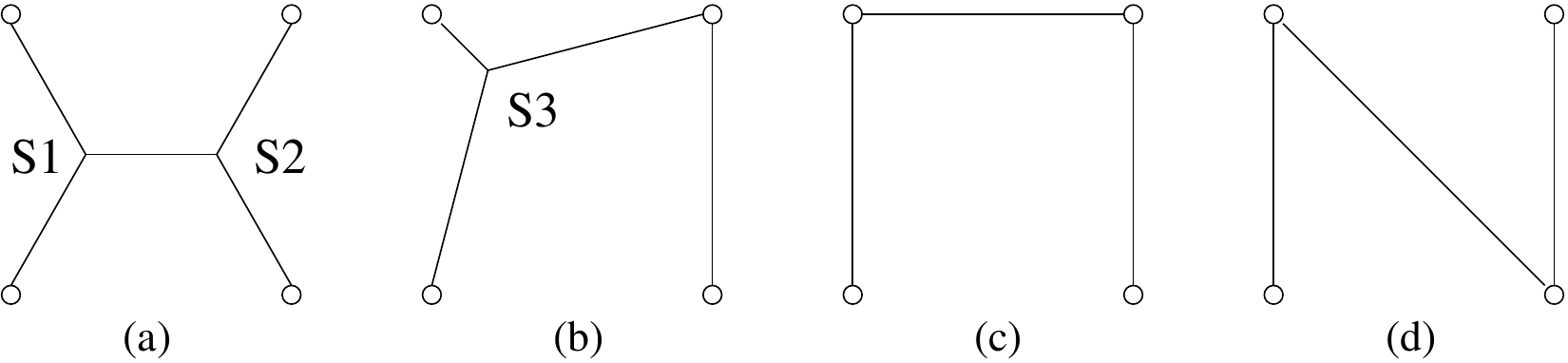}
\caption{Four terminal Steiner and spanning trees.} \label{soapfig1}
\end{figure}

\begin{figure}[h]
\includegraphics[width=6.5in]{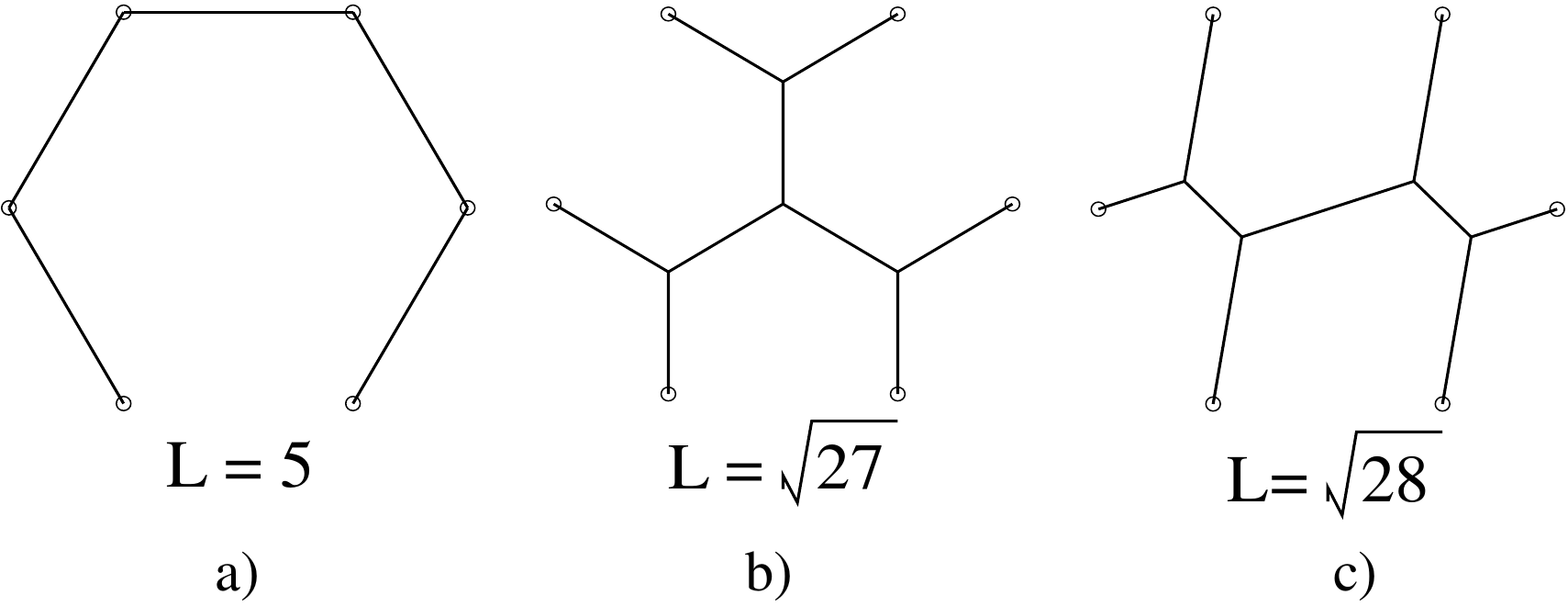}
\includegraphics[width=6.5in]{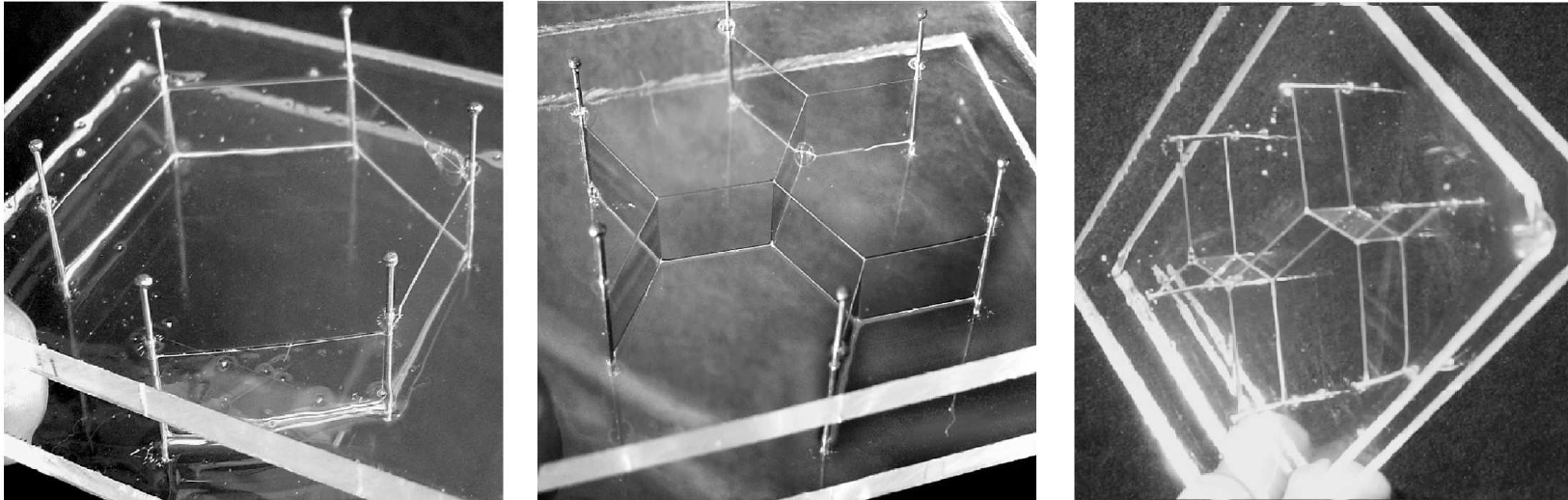}
\caption{Commonly reported soap film configurations for 6-pin regular
  hexagon.}
\label{soapfig2}
\end{figure}

Usually a {\it Steiner tree} in contrast to a spanning tree connects 
the given points
introducing extra point or points ( called {\it Steiner points} or
{\it Steiner vertices}) to reduce the total path length.   
For a given number of initial points there are various possible Steiner tree
configurations. Among these configurations one (or more in some cases) 
will correspond to absolute minimum configuration, whereas the others 
will correspond to local minimum configurations. 
 A Steiner {\it minimal} tree has shortest total path 
length\footnote{There are 
cases (like the present six-pin problem), where the minimal 
tree is a spanning tree(see Fig.\ref{soapfig2}(a)).}. 
There are cases where there are two or more
different configurations have the same path length, we call these
degenerate configurations.
Often the symmetry of the system is responsible for the degeneracy, 
that is one configuration can be identified with the
other degenerate configuration through a simple rotation or reflection
(same as the symmetry of the system). In contrast there are other 
cases where completely different looking configurations are degenerate. 
Degenerate configurations Fig. 8 b) and Fig. 8 c) are examples of the latter 
type. Examples of regular four-terminal trees are
shown in Fig. \ref{soapfig1}. If each side of the square is taken
as one unit, the total length of the stems connecting the four terminals
are 1+$\sqrt{3}$, 1+$\frac{1}{\sqrt{2}}(1+\sqrt{3})$, 3 and
$2+\sqrt{2}$ units,
respectively, in (a), (b), (c) and (d). In this case, the Fig.
\ref{soapfig1}(a) is the Steiner minimal tree. 
The  tree in Fig. \ref{soapfig1}(b) is a Steiner tree, whereas the
trees of Fig. \ref{soapfig1}(c) and (d) are examples of spanning
trees. The new vertices in
the tree (other than initial given points), eg. S1 and S2 in Fig.
\ref{soapfig1} (a) and S3 in Fig. \ref{soapfig1}(b), are the
Steiner points or Steiner vertices. At Steiner point, three branches
of the tree meet and make 120 degree angle with one another.
Moreover, given $n$ initial points, one can have at most $(n-2)$
Steiner points in a Steiner tree. 

The above Steiner problem is also known as Motorway problem,
\emph{i.e.} joining several towns and cities with motorable roads
having the least road length. The motorway problem is of practical
importance because it optimises the cost of constructing roads,
electricity and gas pipe lines etc.  linking different towns and
cities (of course, in these cases the town and cities do not lie
on a plane due to the earth's curvature; and hence, the
corrections due to the curvature should be incorporated in the
problem accordingly).

Alternatively, the Steiner problem can also be studied
experimentally, using soap films connecting $n$-pins between two
parallel plates \cite{Courant,Isenberg,Lovett}. A soap film is
formed between two parallel transparent plates separated by a
small distance and connected by n-pins perpendicular to the plates
(Fig. \ref{soapfig2} for six pins). These $n$-pins act as the given
$n$-terminals of the Steiner problem. When these parallel plates are
immersed in a soap solution and then withdrawn, soap films of
different configurations are formed. The soap film formed, in this
way, would try to achieve the minimum surface energy (local or
absolute) and hence form a stable configuration. The different
ways of drawing the plates out of the soap solution may lead to
different stable configurations. Normal projection of the films on
either plate gives a Steiner tree.

In this present note we study the 6-point (regular) Steiner
problem (that is the given 6-points are the vertices of a regular
hexagon) in detail. We will study the problem with the aid of soap
films following the above mentioned method. We have kept the
length between successive pins as 1 inch and the separation
between plates equal to 1 cm. We have used pins with four different
diameters, viz. (0.32 $\pm$ 0.02)mm, (0.80 $\pm $ 0.02)mm, 
(2.60 $\pm $ 0.02)mm and (4.48 $\pm $ 0.02)mm. 
The soap solution was prepared using 10 ml of liquid detergent and 10 ml
of glycerine in 2 litres of water.

The existing literature always refers to the following three  soap
film configurations \cite{Isenberg,Lovett} (see Fig. \ref{soapfig2}),
which are obtained when one uses very fine needle like pins (in our case we
observed with the pin diameter = (0.32 $\pm$ 0.02)mm. The critical diameter of
the pins, which would observe only the above three configurations
might also depend on the viscosity of the solution used to produce the
films.

 Energies associated with these films are proportional to
their lengths. For configurations (a), (b) and (c) of Fig.
\ref{soapfig2}, lengths, L, of the Steiner trees are found to be equal
to $5$, $\sqrt{27}$ and $\sqrt{28}$, respectively, if the distance
between the neighbouring terminals is taken as 1. Fig. \ref{soapfig2}a)
is the minimal tree for the regular 6-pin problem. Here, we shall
report many new configurations (some of which appeared more
frequently than  Fig. \ref{soapfig2}(b)), observed by us while
repeating the 6-pin soap film experiment using
pins of diameter (0.80 $\pm $ 0.02)mm (not very fine or thick pins). 
The spanning tree configurations
are obtained when thick pins of diameters (2.60 $\pm $ 0.02)mm and
(4.48 $\pm $ 0.02)mm are used on the terminals.
We shall also suggest ways of
constructing these films (i.e the Steiner trees) theoretically.

Section 2 presents the expression for calculating the length of
general 3-pin Steiner tree, which will be treated as the basic
building block for further analysis. Section 3 discusses the
methodology, for creating Steiner trees for general $n$-terminal
Steiner problem. In section 4, we report the new configurations
observed for 6-pin soap films with pin diameter (0.80 $\pm $ 0.02)mm.
 Three of which will be dealt in
detail. We shall present the method of constructing them
geometrically and shall calculate their lengths exactly. 
In section 5 we present the spanning trees using thick pins at terminals.
Finally,
we will catalog all possible configurations (existing as well as
the new ones) for the 6-pin soap film problem having lengths less than
or equal to 6 units(taking neighbouring pin separation as 1 unit). 
Section 6 proposes
an exact two-parameter empirical formula for calculating the lengths of various
6-terminal Steiner trees and spanning trees. 
Section 7, discusses about the loose ends
of the problem and also about the scope of further studies.

\section{Three-pin Steiner minimal tree}

\begin{figure}[h]
\begin{center}
\includegraphics[width=2.8in]{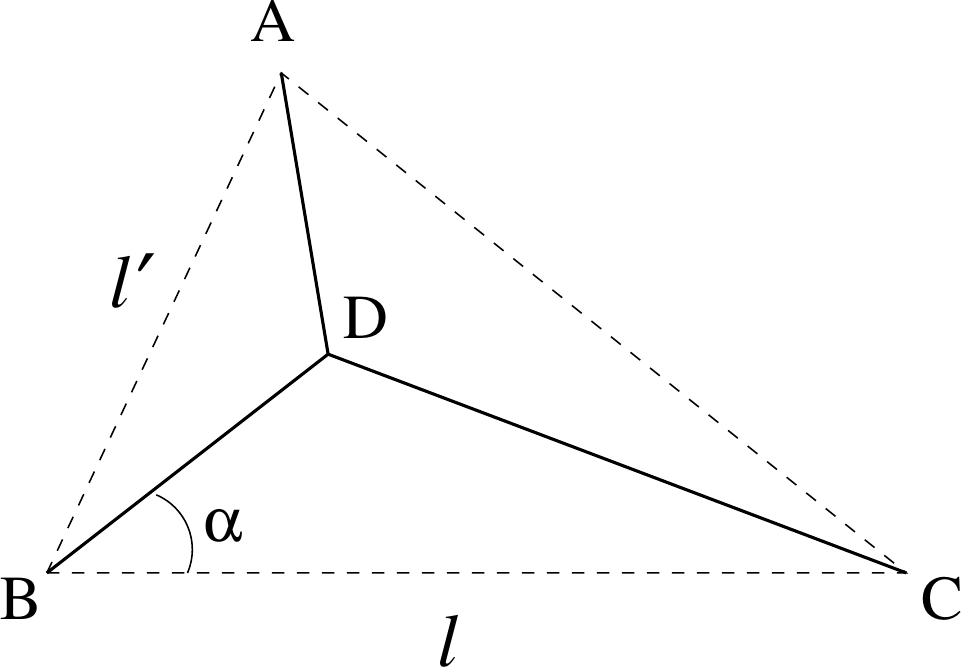}
\caption{Steiner tree (solid lines) in a 3-terminal Steiner
(vertices of the triangle ABC) problem.} \label{soapfig4}
\end{center}
\end{figure}

In this section we give the expression for obtaining the length of
3-terminal Steiner minimal tree. The given points A,B, and C are
taken as the vertices of the triangle ABC (has no angle greater
than 120 degree), which unambiguously defined by the length BC
(=$l$) and AB (=$l^\prime$)and the angle ABC (or simply $\angle$B)
$\leq$ 120$^\circ$ [refer Fig. \ref{soapfig4}]. The Steiner
minimal tree has one Steiner vertex D, such that
$\angle$BDC=$\angle$CDA=$\angle$ADB=120$^\circ$. Hence, the angle
$\angle$DBC=$\alpha$, is given by
\begin{equation}
\tan\alpha=\frac{{\rm {BC}}\sin(\pi/3)-{\rm {AB}}\sin(\pi/3-{\rm
B})}{{\rm {BC}}\cos(\pi/3)+{\rm {AB}}\cos(\pi/3-{\rm B})}
=\frac{l\sin(\pi/3)- l^{\prime}\sin(\pi/3-{\rm
{B}})}{l\cos(\pi/3)+ l^{\prime}\cos(\pi/3-{\rm {B}})}.
\end{equation}
 The length of the Steiner minimal tree, L, is given
by
\begin{eqnarray}
{\rm L}&=&{\rm{AD+BD+CD}} =\frac{1}{\sqrt{3}}[(l-2{l^\prime}\cos
B)\sin\alpha+
(l\sqrt{3}+2{l^\prime}\sin B)\cos\alpha],\nonumber\\
&=&\frac{1}{\sqrt{3}} \frac{1}{\sqrt{1+\tan^{2}\alpha}}
[(l-2{l^\prime}\cos B)\tan\alpha+(l\sqrt{3}+2{l^\prime}\sin B)].
\label{steilength}
\end{eqnarray}
\noindent Note that, if $\angle\rm{B}$ is greater than
120$^\circ$, L=$l+l^\prime$.

\section{Methodology}
In this section we describe, in detail, the geometrical way of
constructing all configurations we came across, while repeating
the 6-pin soap film experiments. To the best of our knowledge,
some of these configurations were never discussed in the
literature before. We shall mention here that the final stable soap film
configuration very much depends on the diameter of the terminal pins
and also on the way the parallel plates are
withdrawn from the soap solution. Usually, a particular way of
withdrawing results in a particular type of configuration. As we
will see that these new configurations could be reduced to
independent 5-pin, 4-pin and 3-pin problems with trivial 1-pin,
2-pin and 3-pin extensions, respectively. This may be one of the
reasons that they were ignored earlier. But we would like to
stress that the reductions make the configurations no way less
interesting than those already discussed. It is interesting to note
here that the Steiner minimal tree in the this problem
(Fig. \ref{soapfig2} a)) can also be regarded as 1-pin trivial extension
of rest 5-pin Steiner minimal tree. As we calculate the
length of these new configurations exactly, we shall find that
they are not arbitrary but are mathematically related in an
interesting way, thanks to the symmetries of the regular hexagon.
Here we suggest a simple method of constructing Steiner trees for
more than 3-terminals, geometrically. This method is particularly
useful if one has a symmetrical arrangement of terminals.  The
procedure will be clear by the following regular 5-terminal,
6-terminal and 8-terminal examples.

We follow the following three steps - (i) First, we join each
terminal with other terminals with straight lines. Basically,
this process triangulates the area inside the terminals. (ii)
Next, we look for non-overlapping \emph{similar} triangles with same
handedness(chirality)  joining  all the terminals (see
 shaded triangles in Fig. \ref{soapfig5}) 
and are themselves linked in special ways with one another. The
linking points of various similar triangles contain
the similar vertices of each of the linked triangles.
(iii) Finally, we construct Steiner
minimal tree for each of these triangles (solid lines in Fig.
\ref{soapfig5}) and that results in a Steiner tree configuration
of the original problem. If there is a single common vertex to two
of these non overlapping triangles, then a soap film passes
through that vertex [Fig. \ref{soapfig5}(a) and (c), point A for
  example]. In this link a reflection about
the common point (A in Fig. \ref{soapfig5}(a) and (c))) and a
scaling  would identify the triangles. The above ensures that a soap
film can pass through the common vertex. Otherwise,
if the vertex is common to three triangles then that vertex will
be a Steiner vertex [point B in Fig. \ref{soapfig5}(b)]. A more intricate
method of triangulation, which uses \emph{non-similar} triangles,
is also considered to obtain the new configurations geometrically.
This method will be illustrated for our 6-pin examples later (new
configuration b) and c) in Section 4. Moreover,  both these methods
are also of great help in calculating the length of a Steiner
tree. Length of the tree is obtained just by adding the
individual tree lengths of each of the triangles using eqn.(\ref{steilength}).

\section{New Steiner trees}
In the following we discuss the new Steiner trees configurations
 observed by us using the pin diameter (0.80$\pm$0.02)mm
(see Fig.\ref{soapfig6}). We shall start with six terminals ABCDEF. 
In what follows we shall
assume the distance between nearest neighbours, AB=BC=CD=DE=EF=FA=1.

\noindent {\bf Configuration  a)} The configuration, shown in the Fig.
\ref{soapfig6}(a), has three Steiner vertices. The geometrical
construction is done using the method described in the previous
section. Three similar triangles, BCG, GDH and HEA join five
terminals (5 vertices of the regular hexagon A, B, C, D and E).
Each of these is a right angled triangle with acute angles
30$^\circ$ and 60$^\circ$. Triangles BCG and GDH have one common
point G and triangles GDH and HEA have a common point H.
 The two similar triangles which have a common
point (say G), are formed by two oblique lines (CH and BD for this
case) intersecting two parallel lines BC and AD. Now drawing
Steiner minimal tree of these triangles individually gives a
Steiner tree for the pentagon ABCDE. This 
construction guarantees that the stems passing through the common
points, G and H, in the individual triangles have the same elevation
(a mandatory condition to have soap films passing through the
points G and H). The sixth terminal F is trivially connected with
the terminal E to complete the configuration. Geometrically, it is
easy to see that the point H divides the diagonal, AD, in the
ratio 1:3, \emph{i.e} DH and HA are 1/2 and 3/2, respectively.
Hence, one can calculate the exact length of the Steiner tree for
this configuration. The length of the configuration is calculated
to be equal to 1+$\sqrt{21}$ using the expression in eqn. (\ref{steilength}) 
for individual triangles.

As we have mentioned earlier that this configuration can be viewed
as 5 terminal (A,B,C,D,E) problem with a  trivial 1-terminal
extension (EF). The tree inside the pentagon ABCDE is definitely a
local minimum of the 5-terminal problem, but mathematically the
question may arise whether with the extension stem, EF, the
resultant tree connecting 6-terminals is a local minimum for the
regular 6-terminal problem or not. Physically, when we study the
6-pin soap film problem, we get this configuration as a stable
one, which is formed once in a while. The only delicate point in
this tree is E. A very slight deformation near E results in the
configuration, shown in Fig. \ref{soapfig2}(c). Whereas, it is
quite stable against any other deformation. The thickness of the
terminal pins may be responsible for the physical stability of the
soap film configurations. 

\noindent{\bf  Configuration b)} 
Our next configuration also has three Steiner
vertices and is shown in Fig. \ref{soapfig6}(b). This is
constructed using a more intricate method, as we are no more
dealing with  similar triangles only (as was the case in {\bf a)}). In
this case, the terminals A,B,C,D, and E are joined by the
triangles BCG,GHA, and HDE. Out of these, the triangles BCG and
GHA are similar. To calculate the exact length
of the Steiner tree, it is crucial to find the point H. The
triangle HDE is not similar to GHA but the point H is so chosen
that the elevations of the stems of the trees through H for both
the triangles are exactly same. To locate H, one may start with DH
as `$x$' and HA as `$2-x$', and may calculate the elevations of the
stems, using the expression for $\tan\alpha$ in eqn. (1), which are
given in terms of `$x$'. Equating the elevations for  both the
triangles DHE and GHA, one obtains $x$. The point H divides the
diagonal DA in 1:2 ratio, ie., DH and HA are 2/3 and 4/3,
respectively. As we have mentioned earlier that with (0.80$\pm$
0.02)mm pin-diameter this one is the most
common configuration, which appears in the 6-pin soap film
experiments. Again, like earlier case this is a 5-pin Steiner tree
with a trivial 1 terminal extension EF. Stability criteria are
also same as that of the previous case. The length of the
configuration is calculated to be equal to 1+$\sqrt{19}$, and that
says why this is more frequently obtained than the earlier one
where the length was 1+$\sqrt{21}$.

\noindent{\bf Configuration  c)}
 The third configuration, shown in
Fig. \ref{soapfig6}(c), has two Steiner vertices.  This is
constructed as  4-terminal (ACDE) problem with two trivial 1-pin
extensions, BC and EF. The quadrilateral  ACDE is joined with the
triangle CGA and DGE (non-similar), such that the point G divides
the diagonal DA in ratio 1:5, \emph{ie.} DG and GA are 1/3 and
5/3, respectively. The point G can be calculated in a similar way,
as H was obtained in the case \textbf{b)}. Again, the point G is so
chosen such that the stem elevations through the G for both the
triangles are same. The length of the tree in this case is
2+$\sqrt{13}$.

\section{Spanning trees}
 
 As mentioned earlier each configuration indicates a local 
minimum in the energy structure of the system. Many a times 
these local minima are
not deep enough and are also not well separated so the slightest 
perturbation would slide a particular 
configuration to a different nearby lower local minimum state.
One of the ways to see the different stable configurations is by  scaling the
system larger (keeping the aspect ratio same) so that the separation 
energy between the neighbouring stable configurations become larger.
Using slightly thicker pins at terminals we are perturbing the 
original system and creating some new local minima corresponding 
to non-minimal Steiner tree configurations. Very thick pins at terminals
perturbs the system violently and make spanning tree 
configurations more favourable. They modify the energy structure of
the system drastically such that the local minima corresponding to 
Steiner trees are lost and new local minimum configurations corresponding
to spanning trees begin to appear.  
In these cases one observes sometimes more than three
films join at a terminal in contrast to earlier Steiner tree cases where
the maximum number of films meeting at a terminal was restricted to two.  
Using thicker pins of diameters
(2.60 $\pm $ 0.02)mm and
(4.48 $\pm $ 0.02)mm
in terminals we started observing the spanning tree 
configurations.
We have restricted ourselves to the cases where the total length of the
spanning tree is less than or equal to 6 units.
Three of the spanning trees are shown in Fig.\ref{thihexa}, below.

\section{Exact Empirical Formula }
All configurations, discussed in previous sections with various pin
diameters for 6-pin soap film
problem, differ either in number of Steiner points in the Steiner
tree or in rotational symmetry of the configuration. For example,
the configuration, shown in Fig. \ref{soapfig2}(b) has 3 fold
rotational symmetry with 4 Steiner points, whereas the
one shown in Fig. \ref{thihexa}(b) has 2 fold symmetry with no
Steiner points. Looking at all configurations, obtained by us for
6-pin soap film problem, it appeared that these two `configuration
parameters', number of Steiner points and the symmetry of the
configuration, involved in formation of Steiner trees in this
particular problem, are following a certain mathematical rhythm.
Involving these two parameters, we propose the following empirical
expression for estimating the exact lengths\footnote{ Exact length of the 
tree of the ideal mathematical problem, where terminals
are ideal and have no finite diameter. In the experimental realisations
discussed, actual physical lengths of the films will differ from these
as one has to incorporate the finite diameter effects at the terminals.}, 
L, of the six-pin trees as,
\begin{equation}
{\rm L}_{n,q}=n+\sqrt{(6-n)^{2}-q(6-n-q)}\hspace{0.15in}:\qquad 
{\rm L}_{n,q}={\rm L}_{n,6-n-q}.\label{eqn3}
\end{equation}
where, $q$ is the symmetry of the configuration. $q=2$ for two-fold
and $q=3$ for three-fold symmetry and $q=1$ for the rest.
We define $n=4-p$,
where $p$ is the number of Steiner points
for a particular configuration. `$n$' can also be interpreted as 
total effective nodal number defined in the following way. A terminal
where $m+1$ stems (in our case films) join has an effective nodal
number $m$. 
In terms of $p$ the length of a configuration is given by,
\begin{equation}
{\rm L}_{p,q}=4-p+\sqrt{(2+p)^{2}-q(2+p-q)}\hspace{0.15in}:\qquad 
{\rm L}_{p,q}={\rm L}_{p,2+p-q}.\label{eqn4}
\end{equation}
In addition, it is interesting to
note that for each value of $p$, there are two configurations
(see Table I). All configurations having a length less than 6 units
(calculated using expression (\ref{eqn3})) are tabulated in Table I.
For the Steiner minimal tree configuration (Fig. \ref{soapfig2}(a), 
configuration 
1 of Table I),
the length 5 is obtained when one uses $p=6$ and $q=3$ or $q=5$
 in expression (\ref{eqn4}). There is no apparent significance of these
values of $p$ and $q$ in this case unlike in other configurations
where they count number of Steiner points and the symmetry.
Another exception is configuration {\bf b)} of section 4(
Configuration 4 of Table I). Here $q$ is 2 and apparently there is no
two-fold symmetry. All configurations are shown in Table I with corresponding
lengths, number of Steiner points and symmetry.
 
\section{Summary}
The present note studies the experimental realisation of ($non-minimal$) 
Steiner trees and spanning trees of regular six-terminal mathematical
problem using soap films by varying pin diameter.
As the diameter of the pins are increased one observes that Steiner
tree configurations of greater lengths start appearing. With very thick
diameter pins one finds formation of spanning trees. In the very thin
pin limit formation of the minimal Steiner tree and two other (next
to minimal) Steiner trees dominates. With  medium pin thickness
Steiner trees with greater lengths appear more easily, while in the
thick pin limit the spanning tree formation is dominant. 
It is also interesting to note that if bubbles are absent in the soap
solution only tree configurations are observed. Any film
configuration with loops are never formed when the plates are gently
drawn out of the bubble-free soap solution. 

One of the loose ends in our formula is that  
one of the predicted Steiner trees with length L=$\sqrt{31}$,
corresponding to n=4, q=1 or 5, obtained from  expression (\ref{eqn3}), is not
observed.    

\begin{figure}[h]
\includegraphics[width=6.5in]{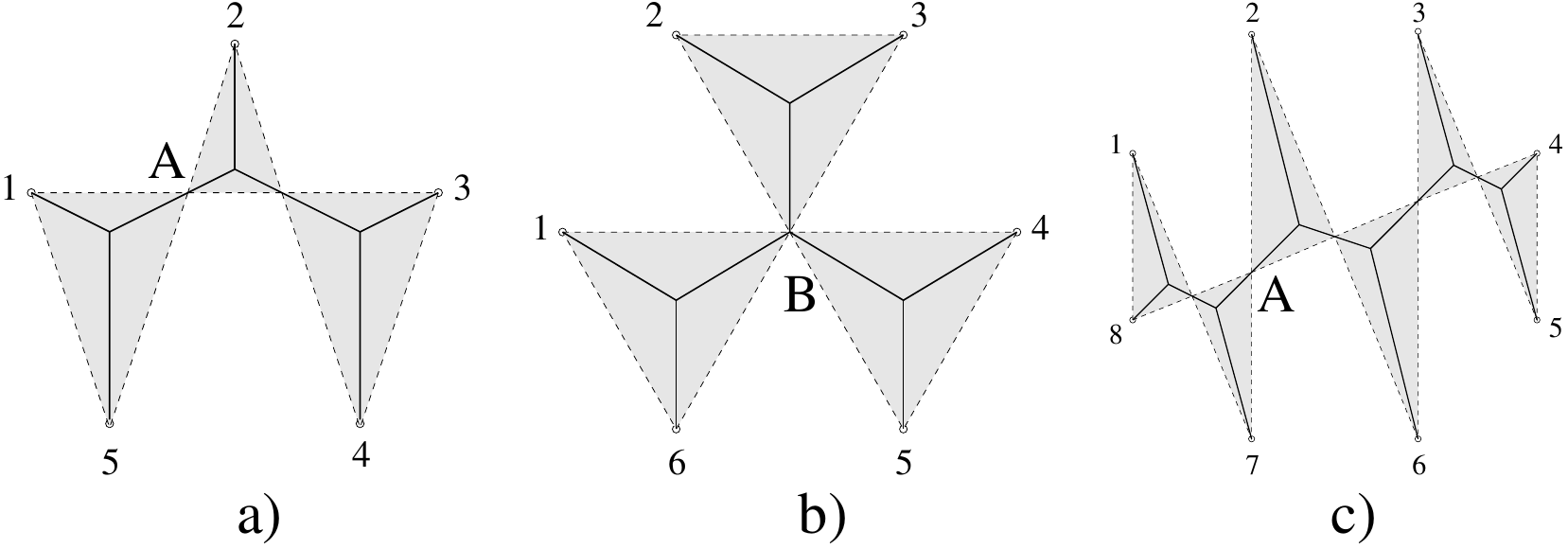}
\caption{Examples of Steiner trees in 5-pin, 6-pin and 8-pin
problem. The triangular shaded areas are chosen to construct the
trees. The solid lines are the  Steiner trees of the original
problem obtained by triangulation technique.} \label{soapfig5}
\end{figure}

\begin{figure}[h]
\includegraphics[width=6.5in]{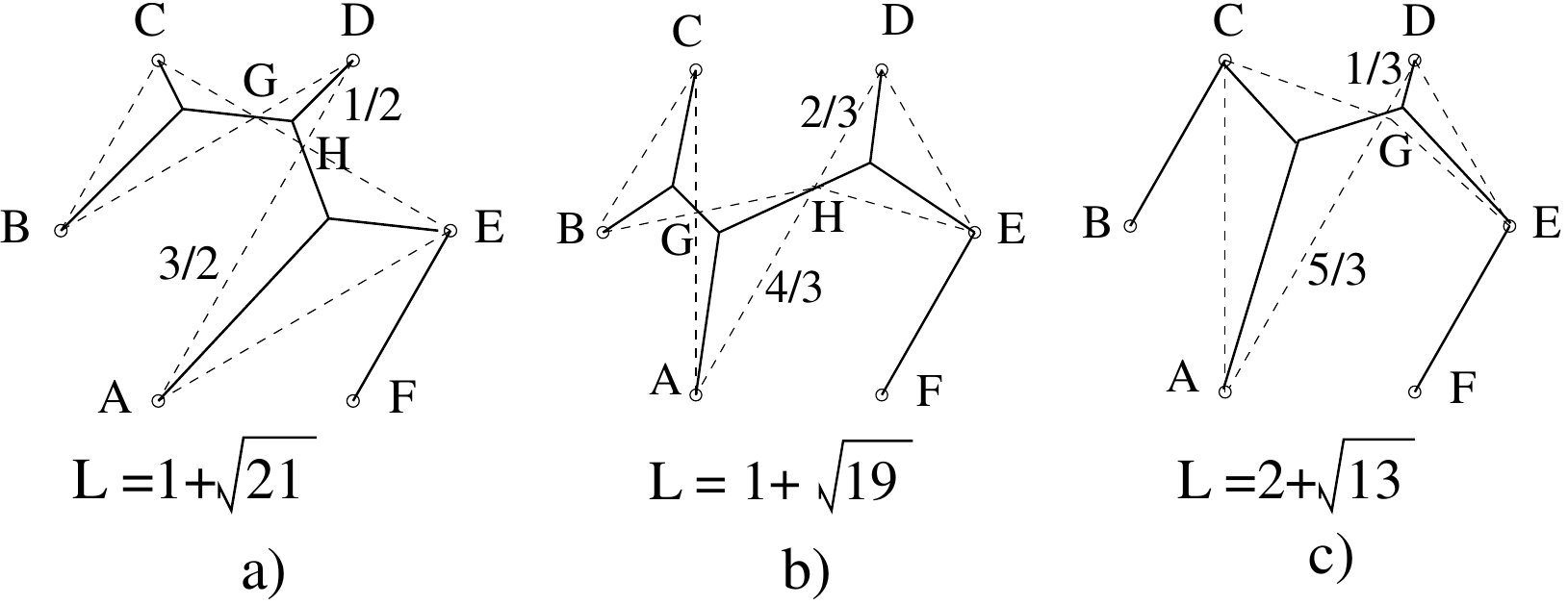}
\includegraphics[width=6.5in]{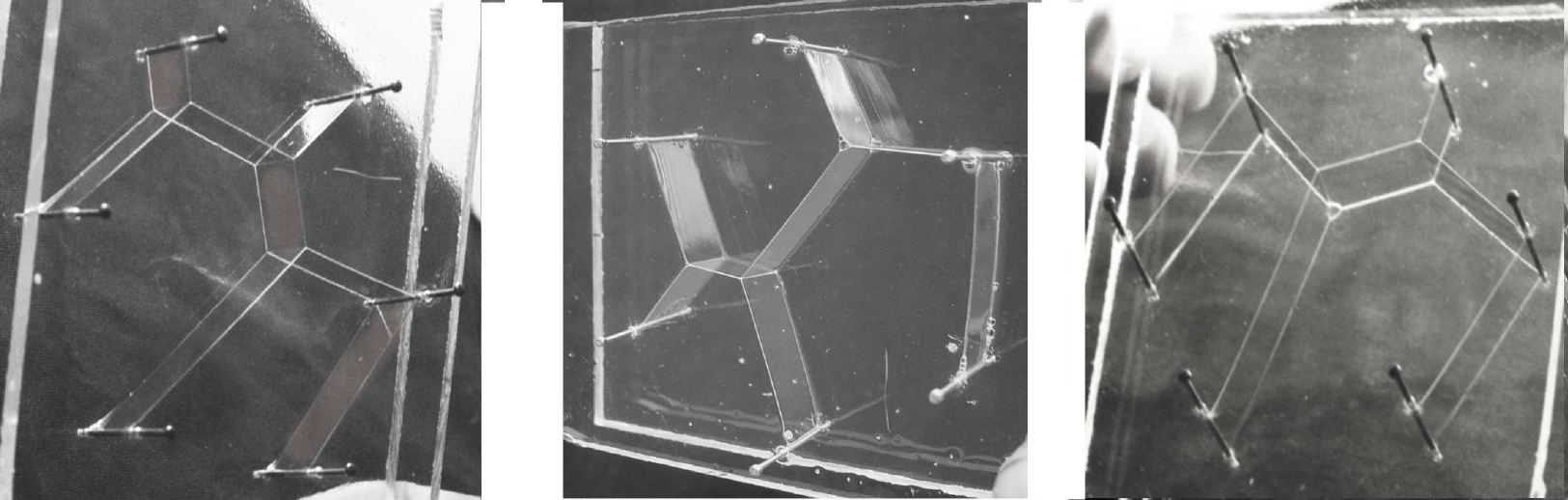}
\caption{New 6-terminal soap films(Steiner trees), discussed in section
4 with pin-diameter (0.80$\pm$0,02)mm.}
\label{soapfig6}
\end{figure}

\begin{figure}[h]
\includegraphics[width=6.5in]{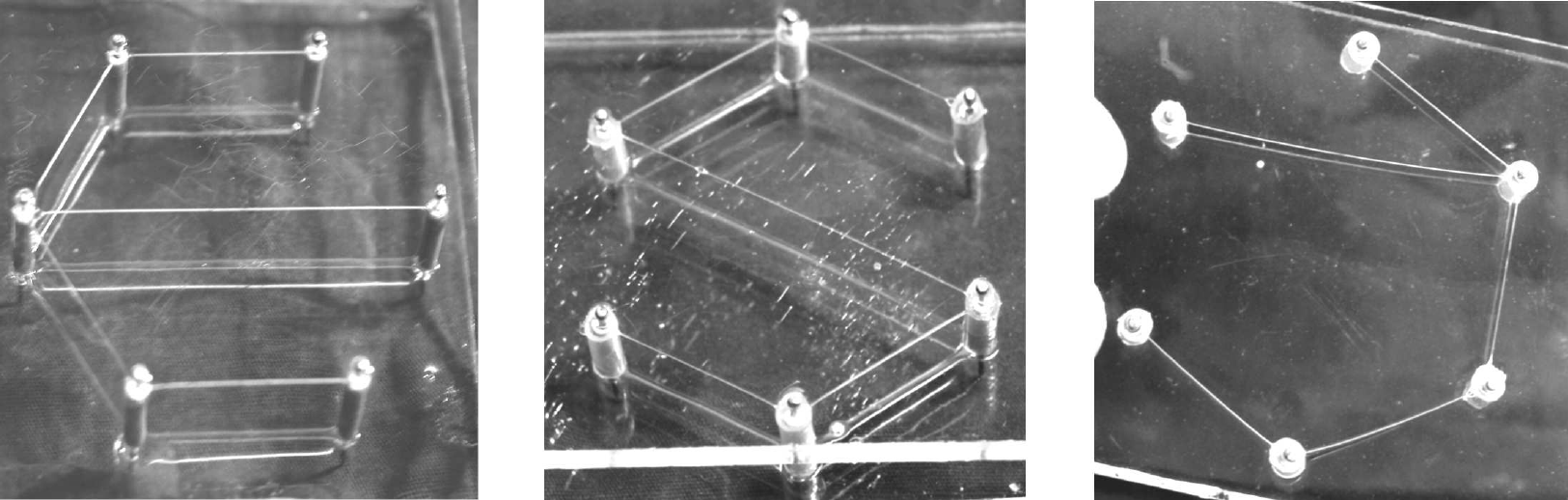} 

\hspace{1.0in} {\Large{\bf a)}}\hspace{2.0in} {\Large{\bf b)}}\hspace{2.0in} 
{\Large{\bf c)}}
\caption{Soap films with thick pins of diameters (2.60 $\pm $ 0.02)mm and
(4.48 $\pm $ 0.02)mm. }
\label{thihexa}
\end{figure}

The same method can be applied to study any $n$-point Steiner
problem. In the Fig. \ref{octa} we show the some of the configurations
(Steiner non-minimal trees) obtained for the regular 8-terminals.
The lengths of these films can also be estimated exactly by the
methodology proposed. For example, the length of the film shown in
Fig. \ref{octa}a) is $(2+\sqrt{2})\sqrt{4+\sqrt{6}}$ (see section 3, Fig.
\ref{soapfig5}c)) with the nearest pin
separation equal to 1 and has 6 Steiner points. 
Configurations in Fig. \ref{octa}b) and c) are degenerate
with 2 Steiner points each and have length equal to $4+{1\over
  2}(2+\sqrt{2})(1+\sqrt{3})$. 

\vspace{0.2in}
\noindent {\Large {\bf Acknowledgments}}
\vspace{0.1in}

\noindent Authors thank Aditi Ghosh for the help rendered during the initial
stages of the work. Authors would also like to thank Anirvan Dasgupta and
Sayan Kar for useful discussions at various stages of the work.


\clearpage

{\bf Figure Captions:}

 Figure 1: Four terminal Steiner and spanning trees.

 Figure 2: Commonly reported soap film configurations for 6-pin regular
  hexagon.

 Figure 3: Steiner tree (solid lines) in a 3-terminal Steiner
(vertices of the triangle ABC) problem.

 Figure 4: Examples of Steiner trees in 5-pin, 6-pin and 8-pin
problem. The triangular shaded areas are chosen to construct the
trees. The solid lines are the  Steiner trees of the original
problem obtained by triangulation technique.

 Figure 5: New 6-terminal soap films(Steiner trees), discussed in section
4 with pin-diameter (0.80$\pm$0,02)mm.

 Figure 6: Soap films with thick pins of diameters (2.60 $\pm $ 0.02)mm and
(4.48 $\pm $ 0.02)mm.

 Figure 7: Table-I:
  6-pin Steiner and spanning trees (* see section 6).

 Figure 8: Soap films observed for the 8-pin system.

\clearpage

\clearpage

\clearpage

\clearpage

\clearpage

\clearpage

\clearpage

\begin{figure}[h]
\begin{center}
\includegraphics[width=6.5in]{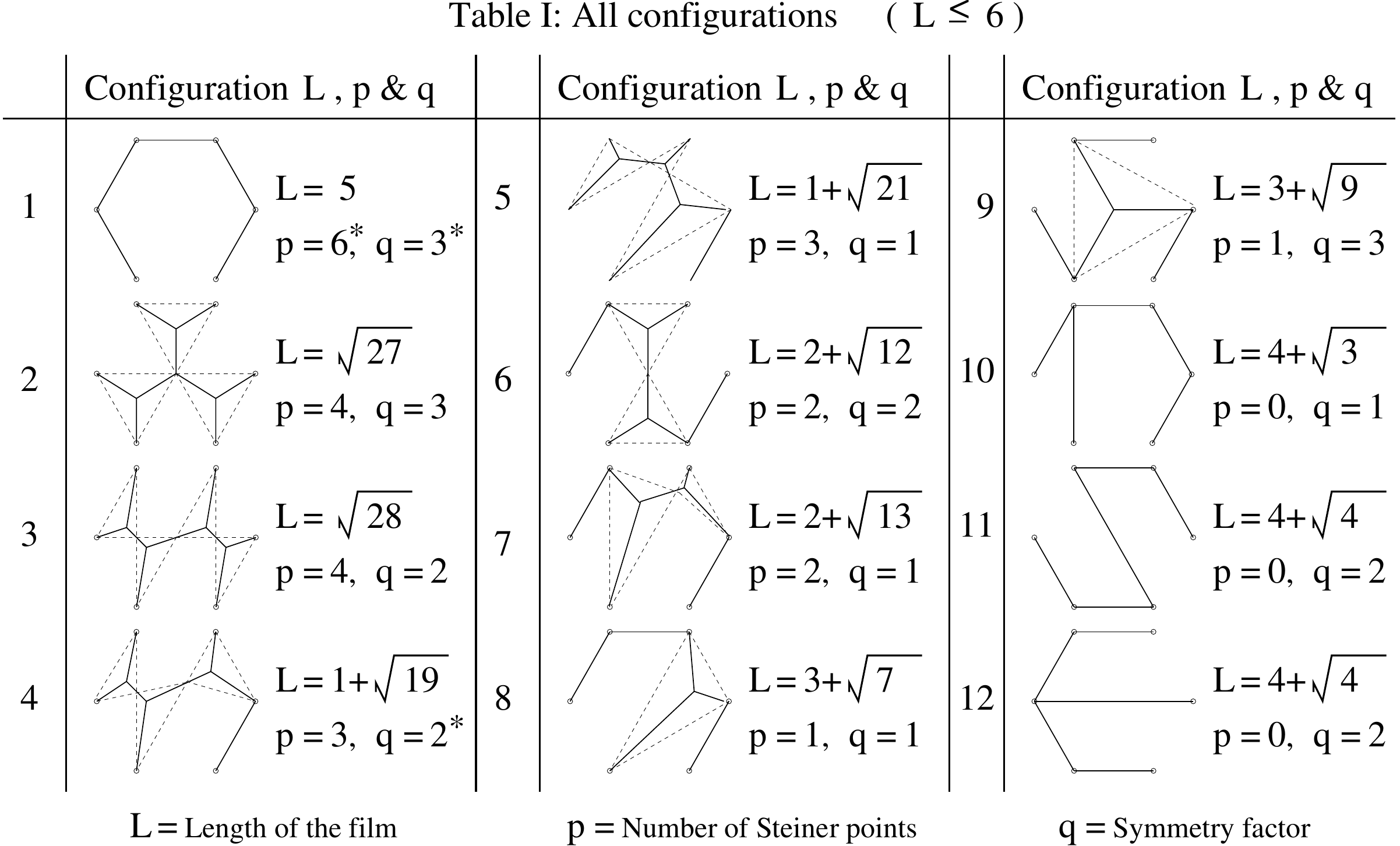}
\caption{Table-I:
  6-pin Steiner and spanning trees (* see section 6).}
\label{hexatab}
\end{center}
\end{figure}

\clearpage

\begin{figure}[h]
\includegraphics[width=6.5in]{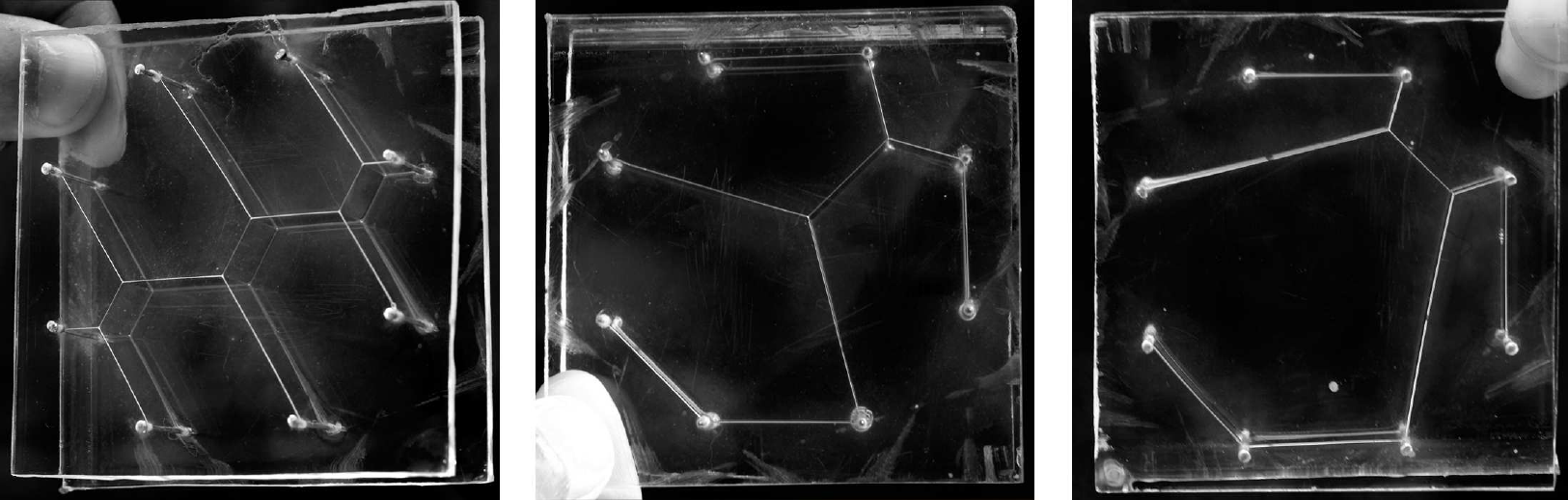} 

\hspace{1.0in} {\Large{\bf a)}}\hspace{2.0in} {\Large{\bf b)}}\hspace{2.0in} 
{\Large{\bf c)}}
\caption{Soap films observed
  for the 8-pin system.}
\label{octa}
\end{figure}

\end{document}